\documentclass[reprint,superscriptaddress,showpacs,amsmath,amssymb,prl]{revtex4-1}

\usepackage{mathtools}
\usepackage{natbib}
\usepackage{siunitx}
\usepackage{array}
\usepackage{xcolor}
\usepackage{subfigure} 
\usepackage{amsmath} 
\usepackage{amsfonts}
\usepackage{amssymb}
\usepackage{graphicx}
\usepackage{hyperref}

\providecommand{\vect}[1]{{\boldsymbol{#1}}}

\DeclareMathOperator{\sech}{sech}

\graphicspath{{../Figures/}}

\begin{document}

\title{Spin-wave driven bidirectional domain wall motion in kagome antiferromagnets}

\author{D.R.~Rodrigues}
\affiliation{Institute of Physics, Johannes Gutenberg-Universit{\"a}t, 55128 Mainz, Germany}
\author{A.~Salimath}
\affiliation{Department of Engineering Sciences, University of Agder, 4879 Grimstad, Norway}
\author{K.~Everschor-Sitte}
\affiliation{Institute of Physics, Johannes Gutenberg-Universit{\"a}t, 55128 Mainz, Germany}
\author{K.M.D.~Hals}
\affiliation{Department of Engineering Sciences, University of Agder, 4879 Grimstad, Norway}
\date{\today}

\begin{abstract}
We predict a mechanism to controllably manipulate domain walls in kagome antiferromagnets via a single linearly polarized spin-wave source. We show by means of atomistic spin dynamics simulations of antiferromagnets with kagome structure that the speed and direction of the domain wall motion can be regulated by only tuning the frequency of the applied spin wave. Starting from microscopics, we establish an effective action and derive the corresponding equations of motion for the spin-wave-driven domain wall. Our analytical calculations reveal that the coupling of two spin-wave modes inside the domain wall explains the frequency-dependent velocity of the spin texture.
Such a highly tunable spin-wave-induced domain wall motion provides a key component toward next-generation fast, energy-efficient, and Joule-heating-free antiferromagnetic insulator devices.
\end{abstract}

\maketitle

Insulator-based spintronics is a promising new pathway for developing the next-generation low-power technologies~\cite{Kajiwara:nature2010,Cornelissen:natphys2015,Lebrun:nature2018}. In contrast to metallic spin electronics, in which the magnetic bits are controlled via the itinerant charge carriers, the spin information in insulators is manipulated by the collective spin-wave (SW) excitations of the magnetic material, i.e., the magnons. This facilitates the transfer and processing of the information with minimal dissipation. In antiferromagnets (AFs)~\cite{Neel:AnnPhys1967,Jungwirth:np2018,Duine:np2018,Gomonay:np2018,Zelezny:np2018,Nemec:np2018,Libor:np2018}, the SW frequencies are in the THz regime, which is a thousand times faster than in ferromagnets. Consequently, besides being more robust against magnetic fields, AFs allow for much higher operational speeds compared to devices consisting of ferromagnetic elements. 

\begin{figure}[ht] 
\centering 
\includegraphics[scale=1.0]{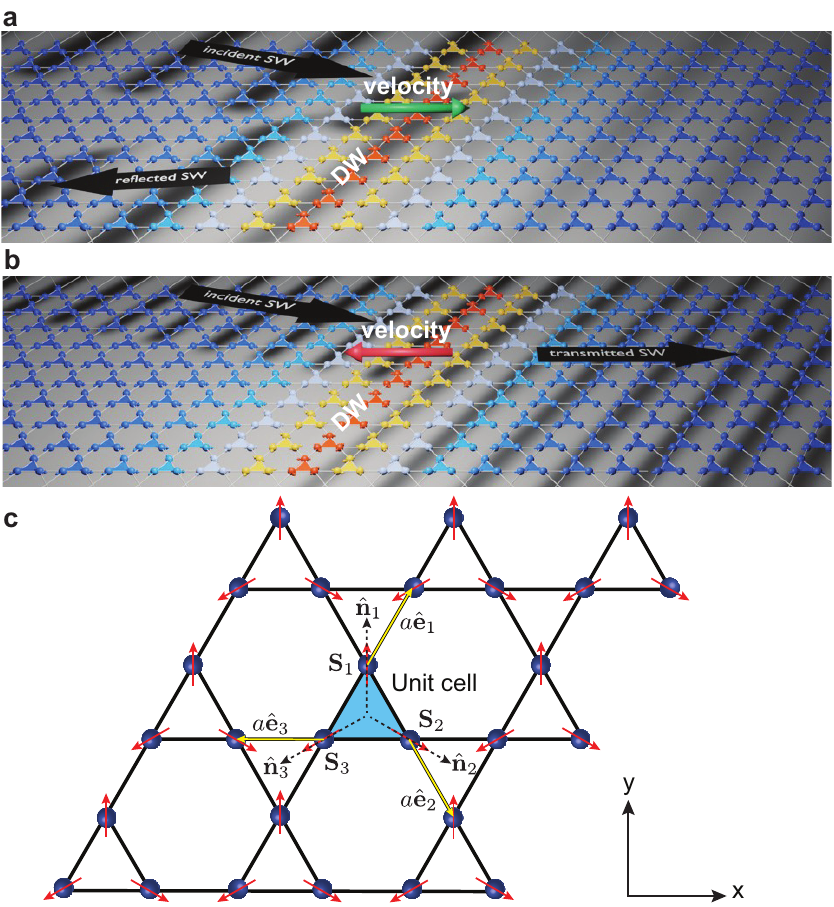}  
\caption{(color online). Frequency controlled SW-induced DW motion. For {\bf a} low ({\bf b} high) frequencies, linearly polarized SWs -- created by an external field along $y$ -- move the DW away from (towards) the SW source. 
The color code is given by $\cos^2\theta$, where $\theta$ is the rotation around the $z$-axis. Thus, in this color scheme, both ground states have the same color.
{\bf c} A kagome AF in one of the two the ground states.}
\label{Fig1} 
\end{figure} 

Such a speed-up by replacing ferromagnets~\cite{Parkin:science2008} with AFs~\cite{Yang:NNano2015, Wornle:nc2019, Back:JPhysD2020} has also been shown for the racetrack memory technology, where domain patterns in nanowires represent the binary data. So far, the study of SW-driven antiferromagnetic domain wall (DW) motion has mainly concentrated on collinear AFs~\cite{Tveten:prl2014,Kim:prb2014,Selzer:prl2016,Yu:prb2018,Oh:prb2019,Shen:apl2020}, which are characterized by a single staggered field. In contrast, non-collinear antiferromagnets (NCAFs) consist of multiple spin sublattices requiring two or three mutually orthogonal order parameter vectors to describe the spin configuration~\cite{Andreev:spu1980}. Due to their complex spin structure, the NCAFs are associated with rich physical phenomena such as Weyl fermions~\cite{Kuroda:nm2017} and a significant spin Hall effect~\cite{Kimata:science2004}. Importantly, a recent experiment demonstrated that it is possible to scan and write DWs in NCAFs using laser pulses~\cite{Reichlova:nc2019}. While theoretical works have predicted that these DWs can be controlled via spin currents in the metallic systems~\cite{Gomonay:prb2012,Zelezny:prl2017}, their coupling to SWs remains unexplored, leaving open several questions of fundamental importance in the context of insulator spintronics.

Here, we theoretically demonstrate that the SW-driven DW motion in NCAFs with kagome structure enables efficient bidirectional DW motion simply by tuning the frequency of the SWs (see Fig.~\ref{Fig1}a-b). 
By developing an effective action of the coupled SW-DW dynamics and performing atomistic Landau-Lifshitz Gilbert (LLG) simulations, we show that certain linearly polarized SWs experience a frequency-dependent DW potential that gives rise to the bidirectional DW motion. Consequently, the placement of the DWs can be manipulated via a single linearly polarized SW-source and thus provides a route toward significantly simplifying the bits' control mechanism in racetrack memories. We expect our predictions to be valid for a large class of materials ranging from iron jarosites~\cite{Matan:prl2006} to Weyl semimetals~\cite{Kuroda:nm2017}.

We model the single-layer homogeneous kagome AF by the spin Hamiltonian
\begin{eqnarray}
H &=& H_{e} + H_{a} + H_{b}, \label{Eq:Hamiltonian}
\end{eqnarray}
where $H_{e}= J\sum_{\langle ij \rangle} \vect{S}_i\cdot\vect{S}_j$ describes the exchange interaction between neighboring lattice spins, $H_{a}=  \sum_{i}  K_z \left( \vect{S}_i\cdot\hat{\vect{z}} \right)^2 - K\left( \vect{S}_i\cdot\hat{\vect{n}}_i \right)^2$ determines  out-of-plane and in-plane anisotropy energies,  and $H_{b}= -\sum_i g\vect{B}\cdot\vect{S}_i$ represents the coupling to an external magnetic field $\vect{B}$. 
Here, $\hat{\vect{n}}_i$ refers to the in-plane easy axis at lattice site $i$, which for the three sublattices of the kagome AF (Fig.~\ref{Fig1}c) is:   
$\hat{\vect{n}}_1 = [0,1,0]$, $\hat{\vect{n}}_2 = [\sqrt{3}/2,-1/2,0]$, and $\hat{\vect{n}}_3 = [-\sqrt{3}/2,-1/2,0]$. Thus, the ground state is given by a $120^{\circ}$ ordering of the sublattice spins such that $\vect{S}_i =  \pm S \hat{\vect{n}}_i$. The unit vectors connecting the three sublattices are 
$\hat{\vect{e}}_1 = [1/2,\sqrt{3}/2,0]$, $\hat{\vect{e}}_2 = [1/2,-\sqrt{3}/2,0]$, and $\hat{\vect{e}}_3 = [-1,0,0]$, whereas $a$ is the lattice constant.  

\begin{figure}[tb] 
\centering 
\includegraphics[scale=1.0]{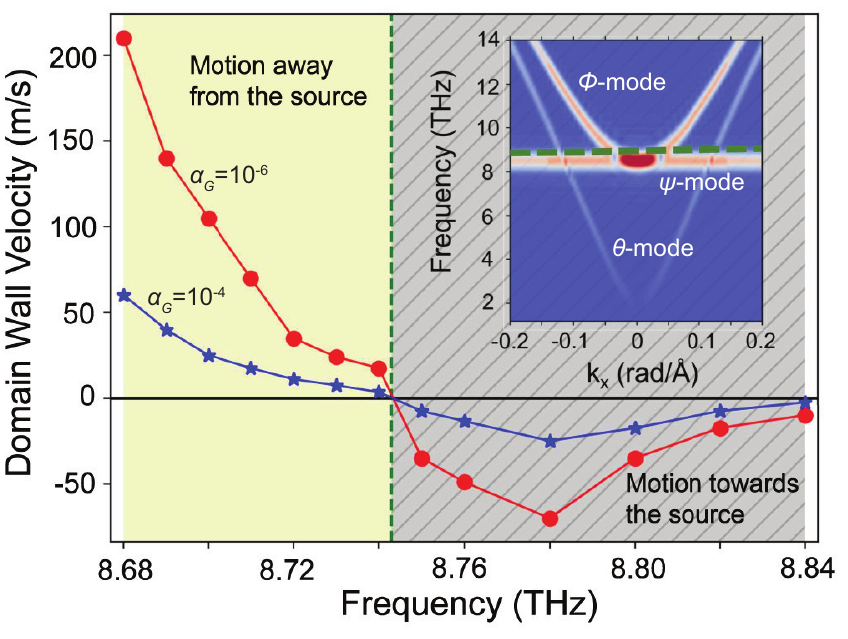}  
\caption{(color online). The DW velocity for different frequencies of the driving field $h_y$. Inset: The SWs' band structure. 
The direction reversal of the motion (green dotted line) occurs  $0.09$~THz above the resonance frequency of the $\psi$-mode.}
\label{Fig2} 
\end{figure}

Fig.~\ref{Fig2} shows the SW driven DW velocity and the SWs' band structure (inset) obtained by atomistic LLG simulations based on Eq.~\eqref{Eq:Hamiltonian}. 
For these simulations we considered a kagome lattice with $a=3$~\AA\  consisting of $500 \times 4$ unit cells, and used material parameters typical for NCAFs~\cite{Sakuma,Szunyogh,Park}: $J= 10.0$~meV, $K_z= 0.9$~meV, $K= 0.03$~meV. The spins are assumed to have $S = 1$ and $g= \hbar \gamma$ with $\gamma= 1.76\times10^{11}$ 1/Ts. 
For the Gilbert damping we consider two different examples, $\alpha_G = 10^{-4}$~\cite{Yu:prb2018} and $\alpha_{G} = 10^{-6}$.  The DW was relaxed at the centre of the nanoribbon with open boundary conditions along $y$ and absorbing boundaries along $x$ to avoid reflection of SWs at the edges~\cite{Comment2}. 
The SWs' band structure was mapped out by applying a ${\rm sinc}$-pulse (i.e., $\vect{B}\sim {\rm sinc} (f t)\hat{\vect{y}}$) and Fourier transforming the resulting response~\cite{Comment3}. 
The SWs were excited by applying a magnetic field, $\vect{B}= B_{0}\sin(\omega t)\hat{\vect{y}}$, where $B_{0} = 80$~mT, locally in a small region near the left edge of the sample. 
The stationary DW velocity was extracted by monitoring the location of the DW center. 
For a more detailed description of the numerics, see Ref.~\cite{Suppl}. The computed DW velocities range up to $200$~m/s, but more importantly, we find that the DW moves away from the source for $\omega < 8.74$~THz, and towards the source for higher frequencies. This change in the direction of the DW velocity at $\omega\sim 8.74$~THz is $0.09$~THz above the resonance frequency $\omega_{\psi 0}\sim 8.65$~THz of the non-dispersive mode (see the inset of Fig.~\ref{Fig2}).

Next, we will develop an effective action description of the coupled SW-DW dynamics. To this end, we write the three sublattice spins of a unit cell as~\cite{Dombre:prb1989,Ulloa:prb2016}
\begin{equation}\label{Eq:Representation}
\vect{S}_{i} (t) = \frac{S \mathbf{R} (t) \left[ \hat{\vect{n}}_{i} + a\vect{L} (t) \right] }{ \| \hat{\vect{n}}_{i} + a\vect{L} (t)  \| }, \ \ i \in \{ 1,2,3 \} .
\end{equation}
Here, the vector $a\vect{L}$ represents a small tilting (i.e., $\| a\vect{L} \| \ll 1$) of the spins away from their equilibrium direction. 
The rotation matrix $\mathbf{R}$ is the order parameter of the NCAF and is assumed to vary smoothly on length scales comparable to the system's exchange length. 
The action of the spin system is $\mathcal{S}= \sum_i \hbar \int {\rm dt} \vect{A} (\vect{S}_i) \cdot\dot{ \vect{S}}_i - \int {\rm dt} H $, where $\dot{ \vect{S}}_i\equiv \partial_t \vect{S}_i$ and $\vect{A}$ is defined via $\vect\nabla \times \vect{A}(\vect{S}_i)   = \vect{S}_i/S$. Expanding the action $\mathcal{S}$ to second order in $a\vect{L}$ and the space-time gradients of $\mathbf{R}$, and integrating out the field $\vect{L}$, leads to the effective action $\mathcal{S}_{\mathrm{eff}}= \int {\rm dt dV} \mathcal{L}_{\mathrm{eff}}$ with 
\begin{eqnarray}
\mathcal{L}_{\mathrm{eff}} &=& \frac{m}{4} {\rm Tr}\left[  \dot{ \mathbf{R}}^T \dot{ \mathbf{R}} \right] + \Lambda_{\alpha\beta m n} \left[ \partial_{\alpha} \mathbf{R}^T  \partial_{\beta} \mathbf{R} \right]_{mn}  -  \label{Eq:Lagrangian} \\ 
& & R_{kl}K_{klmn} R_{mn}  - \frac{1}{2} \epsilon_{mnl}h_m [\mathbf{R}^T \dot{ \mathbf{R} }]_{nl} . \nonumber
\end{eqnarray}
Here, $m= 2\hbar^2/\sqrt{3}J a^2$, $\mathbf{h}= (\hbar g/2J a_c)\mathbf{B}$,  $\Lambda_{\alpha\beta m n}= \tilde{J} [n_{1m}n_{3n}e_{1\alpha}e_{1\beta} + n_{2m}n_{1n}e_{2\alpha}e_{2\beta} + n_{3m}n_{2n}e_{3\alpha}e_{3\beta} ]$,  
$K_{klmn}= \sum_{i=1,2,3} [ \tilde{K}_z n_{i l}n_{in}\delta_{zm} \delta_{zk} -  \tilde{K} n_{i m}n_{in} n_{i k}n_{i l} ]$, where the exchange and anisotropy constants in the continuum limit are related to the microscopic parameters by $\tilde{J}= 4JS^2/\sqrt{3}$, $\tilde{K}_z = K_zS^2/a_c$, and $\tilde{K}= KS^2/a_c$. $\epsilon_{mnl}$ is the Levi-Civita tensor and $a_c= a^2\sqrt{3}/4$ is the area of the unit cell. Throughout, Einstein's summation convention is implied for repeated indices. The dissipative processes are modeled by the functional~\cite{Comment4}
\begin{eqnarray}
\mathcal{G} &=& \frac{\alpha}{4} \int {\rm dt dV}  {\rm Tr}\left[  \dot{ \mathbf{R}}^T \dot{ \mathbf{R} }\right] . \label{Eq:Dissipation} 
\end{eqnarray}

Hereafter, we consider an infinite NCAF ribbon along $x$. The width along $y$ is assumed to be much smaller than the exchange length $\lambda = (\tilde{J}/4 \tilde{K})^{1/2}$, implying that spatial variations along $y$ can be disregarded. A DW texture corresponds to a smooth in-plane rotation $\mathbf{R}_z (\theta_0)$ of the spins along $x$ between two ground state regions where $\vect{S}_i =  S \hat{\vect{n}}_i$ and $\vect{S}_i =  -S \hat{\vect{n}}_i$, respectively (Fig.~\ref{Fig1} a-b).
The  DW profile $\theta_{0} (x)$ is given by~\cite{Comment:DWprofile}
\begin{equation}\label{Eq:DWprofile}
\theta_{0} (x) = 2\arctan\left[ \exp \left( (x-r)/\lambda  \right) \right],
\end{equation}
where the parameter $r$ represents the center of the DW.

Next, we investigate how the DW in Eq.~\eqref{Eq:DWprofile} couples to the SWs of the system. 
To this end, we represent the rotation matrix in terms of nautical angles: 
\begin{equation}
\mathbf{R}= \mathbf{R}_z (\theta_0 + \theta) \mathbf{R}_y (\phi)  \mathbf{R}_x (\psi).
\end{equation}
Here,  $\psi (x,t)$, $\phi (x,t)$, and $\theta (x,t)$ parameterize linearly polarized SWs corresponding to small rotations about the $x$, $y$, and $z$ axis, respectively, on top of the underlying DW profile $\theta_0 (x,t)$. Substituting the above representation of the rotation matrix into Eq.~\eqref{Eq:Lagrangian} and expanding to second order in the SW amplitudes $\psi$, $\phi$, and $\theta$ yield 
\begin{equation}\label{Eq:L}
\mathcal{L}= \mathcal{T} - \mathcal{U}_e - \mathcal{U}_a - \mathcal{U}_b,  
\end{equation}
where $\mathcal{T}$ is the kinetic term
\begin{equation}\label{Eq:T}
\mathcal{T}= \frac{m}{2}\left[  \left( \dot{\theta}_0 + \dot{\theta}  \right)^2 + \dot{\phi}^2 + \dot{\psi}^2  - 2 \dot{\theta}_0\dot{\psi} \phi \right], 
\end{equation}   
$\mathcal{U}_e$ represents the exchange energy      
 \begin{equation}
\mathcal{U}_e= \frac{3\tilde{J}}{4}\left[ \left( \partial_x \phi  \right)^2 + \left( \partial_x \theta_0 + \partial_x \theta  \right)^2 - \left( \partial_x \theta_0  \right)^2 \phi^2 \right], 
\end{equation}  
and $\mathcal{U}_a$ is the anisotropy energy, which can be expressed in terms of the DW-dependent functions $f_0= -3\tilde{K}\cos^2 (\theta_0)$,
$f_1= 3\tilde{K}\sin (2\theta_0)$, $f_2= 3\tilde{K}\cos (2\theta_0)$, $f_3= 3[\tilde{K} + 2\tilde{K}_z + \tilde{K}\cos (2\theta_0)  ]/4$, and $f_4= 3\tilde{K}\sin (2\theta_0)/2$:
 \begin{equation}
\mathcal{U}_a= f_0 + f_1  \theta + f_2  \theta^2 + f_3 [\phi^2 + \psi^2] - f_4 \phi \psi .
\end{equation}  
The term $\mathcal{U}_b$ describes the coupling to a local magnetic field $\vect{h} (x,t)$, which is used to excite the SWs:
\begin{equation}
\mathcal{U}_b=   -h_x\dot{\psi} - h_y\dot{\phi} - h_z\dot{\theta} .
\end{equation}   
Here, we assume that $\vect{h}$ is located far away from the DW center such that we can disregard its direct coupling to the DW.
The dissipation functional in Eq.~\eqref{Eq:Dissipation} becomes 
\begin{equation}\label{Eq:Dissipation2}
\mathcal{G}= \frac{\alpha}{2} \int {\rm dt dV}  \left[  \left( \dot{\theta}_0 + \dot{\theta}  \right)^2 + \dot{\phi}^2 + \dot{\psi}^2  - 2 \dot{\theta}_0\dot{\psi} \phi \right] . 
\end{equation}   

Eqs.~\eqref{Eq:L}-\eqref{Eq:Dissipation2} is the first central analytical result of this Letter and  represent a general theory of the coupled SW-DW dynamics in kagome AFs~\cite{Comment5}. 

In the absence of a DW, i.e.~$\theta_0=0$, there are no coupling terms between the SW modes $\psi$, $\phi$, and $\theta$. 
Their dynamics are determined by the variational 
equations $\delta \mathcal{S}/\delta\vartheta = \delta \mathcal{G}/\delta \dot{\vartheta}$, with $\vartheta \in \{ \psi ,\phi ,\theta \}$, which yield 
$-\ddot{\vartheta} + \eta_{\vartheta}(3\tilde{J}/2m)\partial_x^2\vartheta - \omega_{\vartheta 0}^2 =0 $ with the following SW dispersion relations: 
$\omega_{\vartheta}= (\omega_{\vartheta 0}^2 + \eta_{\vartheta}(3\tilde{J}/2m)k^2)^{1/2}$.
Here, $\omega_{\psi 0}^2= \omega_{\phi 0}^2 = 3(\tilde{K} + \tilde{K}_z)/m$, $\omega_{\theta 0}^2= 6\tilde{K}/m$,  $\eta_{\psi}=0$, and $\eta_{\phi}= \eta_{\theta}= 1$. 
Note that the $\psi$-mode is dispersionless. 

The DW couples the $\psi$ and $\phi$ modes. The $\theta$-mode remains disentangled and its dynamics resembles the situation of the linearly polarized SWs in collinear AFs~\cite{ThetaMode}. 
Therefore, below, we concentrate on the $\psi$ and $\phi$ modes, whose dynamics differ markedly from those of ferromagnets and collinear AFs.
A variation of Eqs. \eqref{Eq:L} and \eqref{Eq:Dissipation2} with respect to $\phi$ and $\psi$ leads to 
\begin{subequations}
\label{Eq:PhiPsi}
\begin{align}
\label{Eq:Phi} 
-\ddot{\phi} -\frac{\alpha}{m}\dot{\phi} &= \left[  -\frac{1}{2m_0}\partial_x^2 +U_{\phi}  \right]\phi + g\psi +  \frac{1}{m}\dot{h}_y, \\
-\ddot{\psi} -\frac{\alpha}{m}\dot{\psi} &=  U_{\psi} \psi + g\phi +  \frac{1}{m}\dot{h}_x, \label{Eq:Psi} 
\end{align} 
\end{subequations}
when $\dot{\theta}_0=0$ and $r=0$. Here, $U_{\phi} (x)= \omega_{\psi 0}^2 - 6U_0\sech^2 (x/\lambda)$, $U_{\psi} (x)= \omega_{\psi 0}^2 - 2U_0\sech^2 (x/\lambda)$, $g(x)= 2U_0 \sech(x/\lambda) \tanh (x/ \lambda)$, $m_0= m/3\tilde{J}$, and $U_0=3\tilde{K}/2m$.
\begin{figure}[tb] 
\centering 
\includegraphics[scale=1.0]{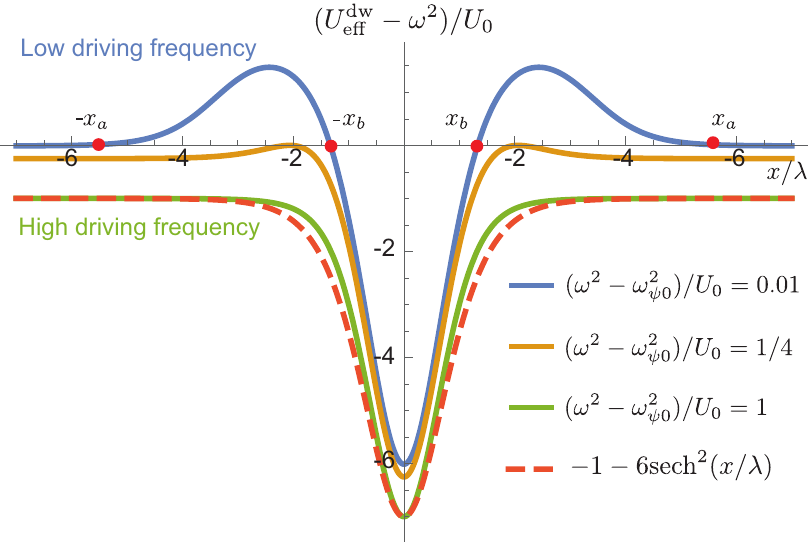}  
\caption{(color online). The effective potential for three different frequencies. For low driving frequencies, characterized by $0< (\omega^2 - \omega_{\psi 0}^2)/U_0 <1/4$, the potential is highly reflective and represents a four turning points (the red dots) scattering problem, whereas at higher frequencies it approaches the P\"oschl-Teller potential (dashed curve).}
\label{Fig3} 
\end{figure} 
We note that the potentials $U_{\phi}$ and $U_{\psi}$ produced by the DW have the form of the reflectionless P\"oschl-Teller potential~\cite{Davies:LTP2015}.
Additionally, the two SW modes interact via the DW-induced coupling terms proportional to $g(x)$.
The coupling $g(x)$ arises from the in-plane anisotropy and in-plane DW corresponding to a rotation of the spins about $z$.
To obtain the stationary solution of Eq.~\eqref{Eq:PhiPsi} we employ the ansatz $[\phi (x,t),\psi (x,t) ]= [\chi_{\phi} (x),\chi_{\psi} (x) ]\exp (-i\omega t)$ and $\vect{h} (x,t)= [h_{0x} (x), h_{0y} (x)] \exp (-i\omega t)$. 
By substituting the ansatz into Eq.~\eqref{Eq:Psi} and \eqref{Eq:Phi}, we find an expression for $\chi_{\psi}$ 
\begin{equation}
 \chi_{\psi} = \frac{g\chi_{\phi}}{\omega^2 - U_{\psi} + i \frac{\alpha\omega}{m}} - i \frac{\omega h_{0x} }{ m\left(\omega^2 - U_{\psi} + i \frac{\alpha\omega}{m}\right) }  , \label{Eq:PsiEff}
\end{equation}
and the following stationary equation for $\chi_{\phi}$  
\begin{equation}
\left( \omega^2 + i \frac{\alpha\omega}{m}\right) \chi_{\phi} = \left[  -\frac{1}{2m_0}\partial_x^2 +U_{\rm eff}^{\rm dw}   \right]\chi_{\phi} - i\frac{\omega}{m}h_{0y} , \label{Eq:PhiEff}
\end{equation} 
where the effective potential produced by the DW is 
\begin{equation}
U_{\rm eff}^{\rm dw} (x) = U_{\phi} (x) + \frac{g(x)^2}{ \omega^2 - U_{\psi} (x) + i \frac{\alpha\omega}{m} } . \label{Eq:UEff}
\end{equation} 
Eq.~\eqref{Eq:UEff} is the second central analytical result of this Letter and shows that the SWs created by a field along $y$ experience a frequency-dependent DW potential.  
The potential $U_{\rm eff}^{\rm dw}$ is shown in Fig.~\ref{Fig3} for different values of $\omega$.
In the high-frequency limit, the last term in Eq.~\eqref{Eq:UEff} can be neglected, and the potential approaches the P\"oschl-Teller potential $U_{\phi}$.
Thus, in this regime, the SWs transmit through the DW with no reflection. This is similar to the case of linearly polarized SWs in collinear AFs~\cite{Tveten:prl2014}.
In contrast, $U_{\rm eff}^{\rm dw}$ becomes highly reflective as the driving frequency of the SWs approaches the resonance frequency $\omega_{\psi 0} $ of the non-dispersive mode.
In this limit, the $\phi$-mode strongly couples to the non-dispersive SW within the DW. 
Effectively, this leads to a frequency-dependent DW potential that peaks at $\omega_{\psi 0} $ and decays as $\omega^{-2}$ away from the resonance. 
Hence, one can tune $U_{\rm eff}^{\rm dw}$ between a highly reflective and reflectionless potential by simply varying the frequency of $h_y$. 
Below, we show that this provides an advantageous switching mechanism between two regimes with highly distinct SW-driven DW motion. 

The equation of motion for the DW can be written as
\begin{equation}
m \ddot{r}  + \alpha \dot{r} = F_{\rm dw} . \label{Eq:GeneralEOM}
\end{equation}
The left hand side of Eq.~\eqref{Eq:GeneralEOM} describes the evolution of the isolated DW system and follows by substituting $\theta_0$ into Eqs.~\eqref{Eq:L}-\eqref{Eq:Dissipation2} and varying with respect to $r$.
The right hand side of Eq.~\eqref{Eq:GeneralEOM} represents the SW-induced force $F_{\rm dw}$ on the DW. The general expression for $F_{\rm dw}$ is complex and strongly depends on the driving frequency. In what follows, we aim to derive simplified expressions for $F_{\rm dw}$ in the high- and low-frequency limits of $U_{\rm eff}^{\rm dw}$.  

First, consider the high-frequency limit.
In this case, the coupling between the $\phi$ and $\psi$ modes is negligible and Eqs.~\eqref{Eq:L}-\eqref{Eq:Dissipation2} can be projected onto to the subspace spanned by $\phi$ and $\theta_0$. Varying the resulting action and dissipation functional with respect to $\theta_0$ yields the coupling term $\sim \tilde{K} \sin (2\theta_0)\phi^2$ between the DW and the $\phi$-mode.
The time variations of the field $\phi$ is much faster than the typical time scale of the moving DW. 
Therefore, we can in the above coupling term replace $\phi^2 (x,t)$ by the time-averaged quantity $\langle\phi^2 (x) \rangle$.
Followed by an integration over $x$, we find the force (where $\zeta = (x-r)/\lambda$)
\begin{equation}
\lim_{\omega \to \infty}  F_{\rm dw}= \frac{9\tilde{K}\lambda}{\pi }\int_{-\infty}^{\infty} {\rm d\zeta} \langle\phi^2 (\zeta) \rangle  \sech (\zeta) \tanh (\zeta) . \label{Eq:FdwHigh}
\end{equation} 
Eq.~\eqref{Eq:GeneralEOM} with the force \eqref{Eq:FdwHigh} is identical to the equation describing the SW-driven DW motion in collinear AFs~\cite{Tveten:prl2014}. For a field $h_y$ located at $x_0$ producing a SW with frequency $\omega$ and wavevector $k$, the DW velocity is $\dot{r}= -[\omega\chi_{\phi}^2 (x_0) (1+3k^2 \lambda^2)/6k ] \exp (-Q |(r-x_0)/\lambda |) $, where $Q= \alpha\omega/9\tilde{K}k\lambda$~\cite{Tveten:prl2014}. Thus, at high frequencies the $\phi$-mode drives the DW towards the SW-source, which is consistent with the atomistic calculations in Fig.~\ref{Fig2}. 

Next, we consider the low-frequency regime in which $\omega$ approaches $\omega_{\psi 0}$ from above.
In this case, a large fraction of the SWs scatter off the DW producing a pressure that moves the spin texture away from the SW source $h_y$. 
In the absence of dissipation, an expression for $F_{\rm dw}$ can be deduced from the energy-momentum conservation law $\partial_{t} T_{x}^{t} + \partial_{x} T_{x}^{x} = 0$, where $T_{\alpha}^{\beta}= \sum_{\vartheta}\left[ \partial \mathcal{L}/\partial(\partial_{\beta} \vartheta) \right] \partial_{\alpha}\vartheta - \delta_{\alpha}^{\beta}\mathcal{L}$ is the energy-momentum tensor.
Here, $\vartheta \in \{ \psi ,\phi , \theta_0 \}$, $\alpha$ and $\beta \in \{ t ,x \}$, and  $\delta_{\alpha}^{\beta}$ is the Kronecker delta.
From the Langrangian in Eq.~\eqref{Eq:L} we find $T_{x}^t=  \sum_{\vartheta} m\dot{\vartheta}\partial_x\vartheta  - m (\dot{\theta}_0\partial_x\psi + \partial_x\theta_0\dot{\psi})\phi $ and
$T_{x}^x= \sum_{\vartheta}\mathcal{J}_{\vartheta} + \mathcal{J}_{\rm c}$. $\mathcal{J}_{\vartheta}$ are the momentum fluxes from the SWs and moving DW: $\mathcal{J}_{\vartheta}= -(m/2)\dot{\vartheta}^2 - \eta_{\vartheta}(3\tilde{J}/4)(\partial_x\vartheta)^2 + K_{\vartheta}$, where $\eta_{\phi}=\eta_{\theta_0}= 1$, $\eta_{\psi}=0$, $K_{\vartheta}= 3(\tilde{K} + \tilde{K}_z)\vartheta^2 / 2$ for $\vartheta\in \{ \phi ,\psi \}$, and $K_{\theta_0}= -3\tilde{K}\cos^2 (\theta_0)$.
The DW-induced coupling terms yield 
$\mathcal{J}_{\rm c}= (4 m \dot{\theta}_0 \dot{\psi}\phi + 3 \tilde{J} (\partial_x\theta_0)^2 \phi^2 - 3 \tilde{K} (1-\cos(2\theta_0)) (\phi^2 + \psi^2) - 6 \tilde{K} \phi \psi \sin ( 2 \theta_0 ) )/4 $. We assume $\dot{r}$ to be small enough to neglect the back action of the moving DW on the SWs~\cite{Comment1}.
Time-averaging the above conservation law and integrating over $x$ leads to
$ F_{\rm dw} = \langle \Delta \mathcal{J}_{\phi} \rangle $, where
$\langle\Delta \mathcal{J}_{\phi}   \rangle= \langle\mathcal{J}_{\phi} (x_2)  \rangle - \langle\mathcal{J}_{\phi} (x_1)  \rangle $ is the differential pressure produced by the scattered SWs. The time-averaged pressure $\langle\mathcal{J}_{\phi} \rangle $ is evaluated at a position
$x_1\rightarrow -\infty$ ($x_2\rightarrow +\infty$) far to the left (right) of the DW. 
To obtain an analytical expression for the differential pressure
$\langle \Delta \mathcal{J}_{\phi} \rangle$, 
we solve Eq.~\eqref{Eq:PhiEff} (without dissipation) using the Wentzel–Kramers–Brillouin (WKB) approximation. 
In the low-frequency regime, $U_{\rm eff}^{\rm dw}$ represents a four-turning points problem (see Fig.~\ref{Fig3}) for which the 
WKB solution is known \cite{Bohm:book,Suppl}. Based on the WKB solution, we find 
\begin{equation}
\lim_{\omega \to \omega_{\psi}}  F_{\rm dw} =  \frac{3\lambda}{4} k^2 \tilde{J} \chi_{\phi}^2 (x_0) \mathcal{R} (\omega)  . \label{Eq:FdwLow}
\end{equation} 
$\mathcal{R}(\omega)$ is the reflection probability of the $\phi$-mode, which depends on $\omega$ via specific form of the potential $U_{\rm eff}^{\rm dw}$:  
\begin{equation}
\mathcal{R}(\omega)= \frac{ (4\Theta^2 - 1/4\Theta^2)^2 \sin^2 (\Omega )}{4 \cos^2 ( \Omega ) +  (4\Theta^2 + 1/4\Theta^2)^2 \sin^2 (\Omega )} .\nonumber \label{Eq:ReflectionProb}
\end{equation}
Here,  $\Omega= \pi/2 -\int_{-x_b}^{x_b} {\rm dx} \sqrt{2m_0(\omega^2 - U_{\rm eff}^{\rm dw})}$ and $\Theta= \exp \left[ \int_{x_b}^{x_a}  {\rm dx} \sqrt{ 2m_0(U_{\rm eff}^{\rm dw} - \omega^2)} \right]$. 
From Eq.~\eqref{Eq:GeneralEOM}, it is clear that the DW velocity becomes $\dot{r}=  \lim_{\omega \to \omega_{\psi 0}}  F_{\rm dw}/\alpha$.

Eqs.~\eqref{Eq:GeneralEOM}-\eqref{Eq:FdwLow} represent the third central analytical result of this Letter and provide a theory for the DW motion driven by $\phi$-SWs in kagome AFs. 
A transition between the low- and high-frequency regimes occurs for the frequency $\omega_c= \sqrt{U_0/4 + \omega_{\psi 0}^2}$ at which the scattering problem in Eq.~\eqref{Eq:PhiEff} no longer has any turning points. 
Thus, we anticipate a motion away from the source for $\omega \in [\omega_{\psi 0}, \omega_c]$, whereas a movement towards the source is predicted for $\omega > \omega_c$. The material parameters used in the LLG simulation yield (for $\alpha_G= 10^{-4}$) $\omega_c - \omega_{\psi 0}\sim 0.10$~THz and a DW velocity of $\dot{r}= 58$~m/s ($\dot{r}= 7$~m/s) in the low (high) frequency regime, which agree well with the results in Fig.~\ref{Fig2}~\cite{Estimate}.

To conclude, we have found that the three SW bands of NCAFs with kagome structure lead to a nonconventional coupling mechanism between DWs and SWs. While one of the SW bands (the $\theta$-mode) corresponds to one of the two  modes typically found in collinear AFs, the behavior of the other two bands is different. They couple inside the DW, and one of them is non-dispersive. The physics of these two bands is crucial for the bidirectional DW motion observed in NCAFs, having no counterpart in collinear AFs. Importantly, the predicted coupling mechanism enables manipulation of the DWs -- their speed and direction -- via a single linearly polarized SW source.

D.R.~Rodrigues and A.~Salimath contributed equally to this work.
The work received funding from the Research Council of Norway via Grant No. 286889 and the DFG via the Emmy Noether project 320163632 and TRR 173 -- 268565370 (project B12).
A.~Salimath acknowledges A.~Abbout for the assistance during the development of the code. D.R.~Rodrigues and K.~Everschor-Sitte acknowledge H.~Gomonay for discussions on NCAFs.


\begin{thebibliography}{99}
\bibitem{Comment2} In practice, the absorbing boundaries are introduced by increasing the local damping.
\bibitem{Comment3} We Fourier transform the $x$-component of $\vect{n} = \vect{L}_1\times\vect{L}_2$, where $\vect{L}_1= (\vect{S}_3-\vect{S}_2)/\sqrt{3}S$ and $\vect{L}_2= (\vect{S}_2+\vect{S}_3-2\vect{S}_1)/3S$.
The inset of Fig.~\ref{Fig3} shows the magnitude of $n_x (k_x,\omega)$.
\bibitem{Comment4} The damping parameter $\alpha$ in Eq.~\eqref{Eq:Dissipation} is related to the Gilbert damping of the individual spins by $\alpha = 3\hbar\alpha_G/a_c$.
\bibitem{Comment:DWprofile}The  DW profile $\theta_{0} (x)$ can be computed by substituting the rotation $\mathbf{R}_z (\theta_0)$ into the Langrangian~\eqref{Eq:Lagrangian}, which leads to the equilibrium equation
$\partial_x^2 \theta_0 - \sin(2\theta_0)/2\lambda^2 = 0$  and the DW profile in Eq.~\eqref{Eq:DWprofile}.
\bibitem{Comment5} Notice that we consider DWs that preserve chirality. Due to the lack of inversion symmetry breaking terms in Eq.~\eqref{Eq:Hamiltonian}, the dynamics analyzed in this manuscript is independent of chirality.
\bibitem{ThetaMode} The equation of motion for $\theta$ is $-\ddot{\theta} - \alpha \dot{\theta}/m = \left[  -\partial_x^2 / 2m_0 +U_{\theta}   \right]\theta + \dot{h}_z/m$
where $U_{\theta} (x)= 4U_0 \left( 1- 2 \sech^2 (x/ \lambda)  \right) $ is the potential experienced by the SW due to the DW.
\end{thebibliography}
\end{document}